\documentclass[fleqn,10pt]{wlscirep}
\usepackage[utf8]{inputenc}
\usepackage[T1]{fontenc}
\usepackage{textcomp}

\title{Localization precision in chromatic multifocal imaging}

\author[1,2,3,4]{M. Junaid Amin}
\author[2]{Sabine Petry}
\author[2,3]{Joshua W. Shaevitz}
\author[4,*]{Haw Yang}

\affil[1]{Department of Molecular Biology, Princeton University, Princeton, New Jersey 08544, USA}
\affil[2]{Department of Physics, Princeton University, Princeton, New Jersey 08544, USA}
\affil[3]{Lewis-Sigler Institute for Integrative Genomics, Princeton University, Princeton, New Jersey 08544,USA}
\affil[4]{Department of Chemistry, Princeton University, Princeton, New Jersey 08544, USA}

\affil[*]{Corresponding author: hawyang@princeton.com}

\begin{abstract}
	Multifocal microscopy affords fast acquisition of microscopic 3D images. This is made possible using a multifocal grating optic, however this induces chromatic dispersion effects into the point spread function impacting image quality and single-molecule localization precision. To minimize this effect, researchers use narrow-band emission filters. However, the choice of optimal emission filter bandwidth in such systems is, thus far, unclear. This work presents a theoretical framework to investigate how the localization precision of a point emitter is affected by the emission filter bandwidth. We calculate the Cramér-Rao lower bound for the 3D position of a single emitter imaged using a chromatic multifocal microscope. Results show that the localization precision improves with broader emission filter bandwidth due to increased photon throughput, despite a larger chromatic dispersion. This study provides a framework for optimally designing chromatic multifocal optics and serves as a theoretical foundation for interpretting results.
\end{abstract}

\begin{document}
	
	\maketitle
	
	\section{Introduction}
	Fast three dimensional (3D) imaging is crucial for investigating dynamical biological and material processes occurring at sub-second timescales. Multifocal microscopy is a promising widefield imaging and localization tool for probing such processes due to its high-speed 3D imaging ability over large fields of view to allow simultaneous localization of multiple particles spread over a large 3D volume \cite{dalgarno2010multiplane,abrahamsson2013fast}. These instruments typically use a multifocal grating in the emission path. This grating splits the emission light into multiple diffraction orders that are imaged in a staggered manner on a single camera sensor to produce a stack of subimages, each conjugate to a different object plane. This enables imaging speeds of hundreds of volumes per second.

	Multifocal microscopy is particularly suited for single molecule/particle localization applications. In addition to high imaging speed, a key feature of using multifocal microscopy is the long axial acquisition distance range. For example, a 9-plane multifocal microscope with an object plane separation of 1.5 $\mu$m provides approximately 12 $\mu$m axial working depth range. This is in contrast to other high speed localization techniques using PSF engineering which are limited to a < 3 $\mu$m axial range \cite{von2017three}. More recently, localization algorithms utilizing deep learning have been demonstrated for localization over ~4 $\mu$m axial range \cite{nehme2020deepstorm3d} which is still relatively small compared to multifocal microscopy capabilities. Other methods including optical sectioning techniques such as confocal are too slow (typically < 1 volume per second) compared to multifocal imaging which is capable of hundreds of volumes per second. Therefore, multifocal microscopy offers desirable speed as well high axial depth range suited for localization applications.
	
	Due to the wavelength spread of the sample-emitted/scattered light, the multifocal grating in typical multifocal microscopes induce wavelength dependent dispersion effects on the point spread function (PSF) eventually imaged on the camera sensor. Although use of chromatic correction optics minimizes this dispersion \cite{abrahamsson2013fast}, these custom designed optics which include multiple blazed gratings and a multi-faceted prism are difficult to acquire and inaccessible to most researchers. Without chromatic corrective optics, researchers deploy narrow-band emission filters to limit deterioration of the PSF due to the multifocal grating induced chromatic dispersion \cite{ma2016three, yoo2018bayesian, walker2019high, lin2019multi, aminUniform2020PLOSONE}. For such chromatic multifocal microscopes, it is far from clear how to choose the bandwidth of the emission filters. For single molecule localization applications in particular, the relationship between the localization precision and the emission filter bandwidth in a chromatic multifocal microscope has not been formally investigated yet.
	
	In this paper, we derive Cram\'er-Rao lower bounds \cite{cover2012elements} (CRLB) for the 3D position of a single molecule imaged under a chromatic multifocal microscope, and use the resulting analytical expressions to numerically explore the effect of emission filter bandwidth on the 3D localization precision in such systems. The CRLB expressions for a particle’s position for both conventional microscopes \cite{ramHow2005} and multifocal microscopes based on multiple cameras that lack chromatic dispersion \cite{tahmasbi2014designing} have been reported previously. Here, we adapt the approach in \cite{tahmasbi2014designing} and formulate CRLB expressions for a chromatic multifocal microscope.
	
	
	\section{CRLB theoretical framework}
	\subsection{Conventional widefield microscope}
	We start with sketching the theoretical background using a conventional widefield microscope. Consider a point emitter in the object space having a lateral center position $(x_{0}, y_{0})$ located axially a distance $z_{0}$ from the microscope focal plane. Let the function $q_{z0}(x,y)$ describe the image of the emitter on a detector plane at unit magnification, with $(x,y)$ denoting the detector coordinates. Here, $q_{z0}(x,y)$ is represented by the Born and Wolf \cite{born2013principles} expression for the PSF,
	\begin{equation}\label{eq:qz0}
	q_{z_{0}}(x,y)=\dfrac{1}{C_{z_{0}}}\left[U_{z_{0}}^{2}\left(x,y\right)+V_{z_{0}}^{2}\left(x,y\right)\right].
	\end{equation}
	$U_{z_{0}}(x,y)$ in \eqref{eq:qz0} is expressed as,
	$U_{z_{0}}(x,y)=\int_{0}^{1}J_{0}\left(2{\pi}(\mathit{NA})\lambda^{-1}\rho \sqrt{x^{2}+y^{2}}\right)\mathrm{cos}(W(\rho,z_0))\rho\mathrm{d}\rho$,
	where $\lambda$ represents the wavelength, $\mathit{NA}$ is the system numerical aperture, $J_{0}$ is the zeroth order Bessel function of the first kind and $\rho$ is a dummy variable for integration. $W(\rho,z_0)$ is a phase aberration term given by $W(\rho,z_0)=\pi \rho^2 (\mathit{NA})^2 z_0 / \lambda n_\textrm{oil}$,
	where $n_\textrm{oil}$ is the refractive index of the immersion oil. $V_{z_{0}}(x,y)$ in \eqref{eq:qz0} is given by,
	$V_{z_{0}}(x,y)=\int_{0}^{1}J_{0}\left(2{\pi}(\mathit{NA})\lambda^{-1}\rho \sqrt{x^{2}+y^{2}}\right)\mathrm{sin}(W(\rho,z_0))\rho\mathrm{d}\rho$, 
	whereas the normalization term $C_{z_{0}}$ in \eqref{eq:qz0} is formulated as,
	$C_{z_{0}}=\int_{\Re^2}\left[U_{z_{0}}^{2}\left(x,y\right) +V_{z_{0}}^{2}\left(x,y\right)\right]\mathrm{d}x\mathrm{d}y$. 
	
	Continuing, let $N$ be the mean total number of photons striking the full detector area in an exposure time~$t$. The number of signal photons striking a given pixel area $C_{k}$ of the detector in time $t$ is independently Poisson distributed with mean $\mu_{\theta}(k,t)$, where $k$ is the pixel number. This is given by \cite{ramResolution2007},
	$\mu_{\theta}(k,t)=N\left[M^{-2}\int_{C_{k}}q_{z0}\left(x/M-x_{0},y/M-y_{0}\right)dxdy\right]$,
	where  $\theta=(x_{0}, y_0, z_0)$ denotes the parameter to be estimated and $M$ is the lateral magnification. Note that the argument of $q_{z_{0}}$ in the above equation accounts for scaling due to $M\neq1$ and offset due to non-zero center location $(x_{0}, y_0)$ of the emitter. Given the Poisson nature of this process, the Fisher information matrix element for a pixelated detector is expressed as \cite{snyder2012random},
	\begin{equation}\label{eq:Fisher_no_gaussian_noise}
	I_{ij}=\sum_{k=1}^{P}\dfrac{1}{\mu_{\theta}(k,t)}\left(\dfrac{\partial\mu_{\theta}(k,t)}{\partial\theta_i} \right)\left(\dfrac{\partial\mu_{\theta}(k,t)}{\partial\theta_j} \right),
	\end{equation}
	where $P$ is the total number of pixels in the detector. Let $\mathbf{I}$ be the  Fisher information matrix whose elements are $I_{ij}$, and  $\mathbf{T}$ be the inverse of $\mathbf{I}$.  The Cram\'er-Rao lower bound for a parameter $\theta_i$ is simply,
		\begin{equation}
		\sigma(\theta_i)\equiv T_{ii}^{-1/2}, \label{eq:CRLB}
		\end{equation}
		where $T_{ii}$ are the elements of $\mathbf{T}$. Therefore, the image function $q_{z0}$ in \eqref{eq:qz0} maps the photon distribution of an emitter (located in object space at $(x_0,y_0,z_0)$) onto a detector plane $(x,y)$ and is used to express the number of photons striking each detector pixel $k$ in time $t$ as a Poisson process with mean $\mu_{\theta}(k,t)$. Subsequently, the Fisher information matrix expression for this Poisson process in \eqref{eq:Fisher_no_gaussian_noise} summing over all $P$ detector pixels is used to compute the CRLB for parameters $(x_0,y_0,z_0)$. Note that in practice, when the off-diagnoal terms are very small compared to the diagonal elements, $\sigma(\theta_i)$ can be computed  as $\sigma(\theta_i)=I_{ii}^{-1/2}, $ which produces indistinguishable numerical results (yet computationally much faster). Also note that \eqref{eq:Fisher_no_gaussian_noise} assumes absence of any noise in the detection process. The Fisher information matrix for a noisy image detection process is given in the Appendix, which comprises $\beta(k,t)$ the mean background photon count in time $t$ at the $k^\textrm{th}$ pixel, and $\eta_{k}$ and $\sigma_{\omega,k}$ which denote the mean and standard deviation of the readout noise at the $k^\textrm{th}$ pixel, respectively.
	
	
	\begin{figure}[th!]
		\centering
		\includegraphics{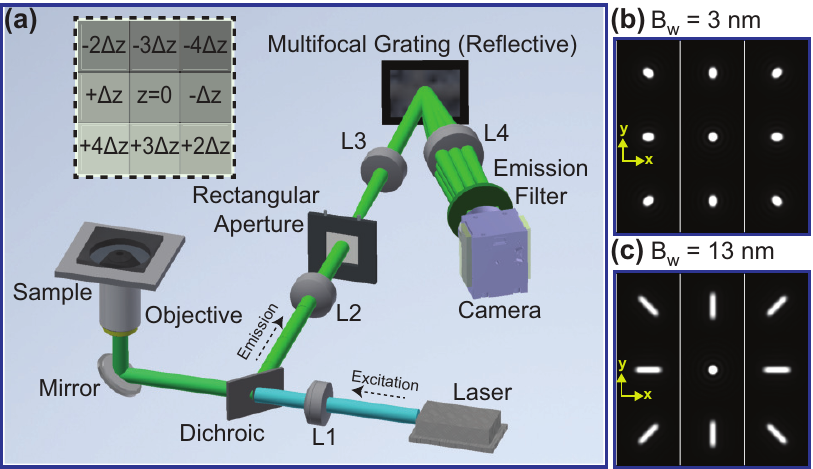}
		\caption{(a) Optical diagram of typical multifocal microscope. L$_1$ focuses the laser onto the objective back focal plane. The rectangular aperture placed at primary image plane, formed by L$_2$, controls the imaging field of view to avoid overlapping subimages at the camera. L$_3$ forms a Fourier transform of the primary image at the multifocal grating. Diffractive patterns on the multifocal grating enable multiple diffraction orders which are captured by lens L$_4$ and imaged onto a camera. Inset shows subimage arrangement on camera sensor with each subimage labeled with its conjugate object plane location relative to the $z=0$ plane, (b--c) Simulated multifocal images of a point source at $\Delta z=0$\,nm showing chromatic dispersion on the PSF at different bandwidths ($B_{\rm w}$).}
		\label{fig:myfig1}\hrule
	\end{figure}
	\subsection{CRLB for single emission wavelength multifocal microscope}
	We next generalize the above results to monochromatic multifocal imaging. To make the theoretical discussion more concrete, Fig.~\ref{fig:myfig1}a shows a representative optical layout for a 9-plane configuration, where $\Delta z$ is the plane separation in object space. Let $\mathbf{I}_\textrm{mfm}^\textrm{mon}$ be the Fisher information matrix for a multifocal microscope having monochromatic wavelength emission. To compute  $\mathbf{I}_\textrm{mfm}^\textrm{mon}$, one needs to calculate the Fisher information matrices for each of the total $N_\textrm{s}$ subimages and sum them up since Fisher information is additive. This is expressed as, $\mathbf{I}_\textrm{mfm}^\textrm{mon}=\sum_{j=1}^{N_\textrm{s}}\mathbf{I}_{j}$, where $\mathbf{I}_{j}$ represents $\mathbf{I}$ computed for subimage $j$ with $j=1, 2, ..., N_\textrm{s}$. Each $\mathbf{I}_{j}$ is computed via Eqs.~(\ref{eq:qz0}) and (\ref{eq:Fisher_no_gaussian_noise}) using the appropriate $z_{0}$ values for each plane.
	
	\subsection{CRLB for chromatic multifocal microscope}
	In a multifocal microscope with a multi-wavelength emission bandwidth, the multifocal grating induces chromatic dispersion which affects the PSF in the multifocal subimages. Simulated multifocal images of a point source in the object plane focused at the $z = 0$ plane for emission filter bandwidth values of 3\,nm and 13\,nm are shown in Fig.~\ref{fig:myfig1}b and Fig.~\ref{fig:myfig1}c, respectively. These images illustrate how the spreading increases with increased emission filter bandwidth while the PSF spread varies across the subimages, due to the direction of the grating lines forming each of these diffraction orders. Figs.~\ref{fig:myfig1}b and ~\ref{fig:myfig1}c are obtained using $\Delta z= 0$\,nm for clear visual comparison between the different subimages by keeping them at the same focus plane.
	
	For such chromatic multifocal microscopes, the monochromatic $q_{z0}$ no longer represents the image function. Therefore, a new image function $q_{z0}^\textrm{chr}$ which accounts for chromatic dispersion is formulated by summing up the $q_{z0}$ expressions over the wavelengths of operation. First, the diffraction angle $\phi$ for a given $\lambda$, multifocal grating period $d_{\textrm{g}}$ and diffraction order $m$ is found using the grating equation, $\phi=\sin^{-1}\left(m\lambda /d_{\textrm{g}}\right)$  \cite{born2013principles}. Note that chromatic dispersion only affects non-zero diffraction orders. Let the emission filter have a central wavelength $\lambda_\textrm{c}$ and a bandwidth of $B_\textrm{w}$ such that the filter is fully transparent for the wavelength range $\lambda_\textrm{c}-B_\textrm{w}/2$ to $\lambda_\textrm{c}+B_\textrm{w}/2$. For a given $\lambda$ within the emission filter's transmission range, the relative distance on the detector $\delta d$ from the striking location of $\lambda_\textrm{c}$ can be found using the grating equation and trigonometry,
	\begin{equation}\label{eq:lambda_location}
	\delta d(\lambda) = f_4\left[\tan\left(\sin^{-1}\left[\dfrac{m\lambda}{d_{\textrm{g}}}\right]\right)-\tan\left(\sin^{-1}\left[\dfrac{m\lambda_\textrm{c}}{d_{\textrm{g}}}\right]\right)\right],
	\end{equation}
	where $f_4$ is the focal length of $L_4$ (Fig.~\ref{fig:myfig1}a). Assuming that when ${\lambda=\lambda_\textrm{c}}$, the image is located at the center of our chosen detector region $(x=0,y=0)$, $q_{z0}^\textrm{chr}$ can be found by summing over all the spatially shifted versions of $q_{z0}$,
	$q_{z0}^\textrm{chr}=\sum_{i}q_{z0}\left(x/M-x_{0}-\delta d_x(\lambda_i)/M,y/M-y_{0}-\delta d_y(\lambda_i)/M\right)$,
	where $i$ denotes all values of $\lambda$, $\delta d_x$ and $\delta d_y$ are $\delta d$ values in $x$ and $y$ detector coordinates, respectively. This distinction in $\delta d_x$ and $\delta d_y$ accounts for the varying subimage chromatic dispersions (Figs.~\ref{fig:myfig1}b and \ref{fig:myfig1}c). $\mathbf{I}_\textrm{mfm}^\textrm{chr}$, the Fisher information matrix for a chromatic multifocal microscope, can now be formulated as,
	$\mathbf{I}_\textrm{mfm}^\textrm{chr}=\sum_{j=1}^{N_\textrm{s}}\mathbf{I}_{j}^\textrm{chr}$,
	where $\mathbf{I}_{j}^\textrm{chr}$ represents $\mathbf{I}_{j}$ calculated using Eqs. (\ref{eq:qz0}) and (\ref{eq:Fisher_no_gaussian_noise}) where $q_{z0}$ is replaced by $q_{z0}^\textrm{chr}$.
	
	\section{Results}
\begin{figure*}[ht!]
	\centering
	\includegraphics{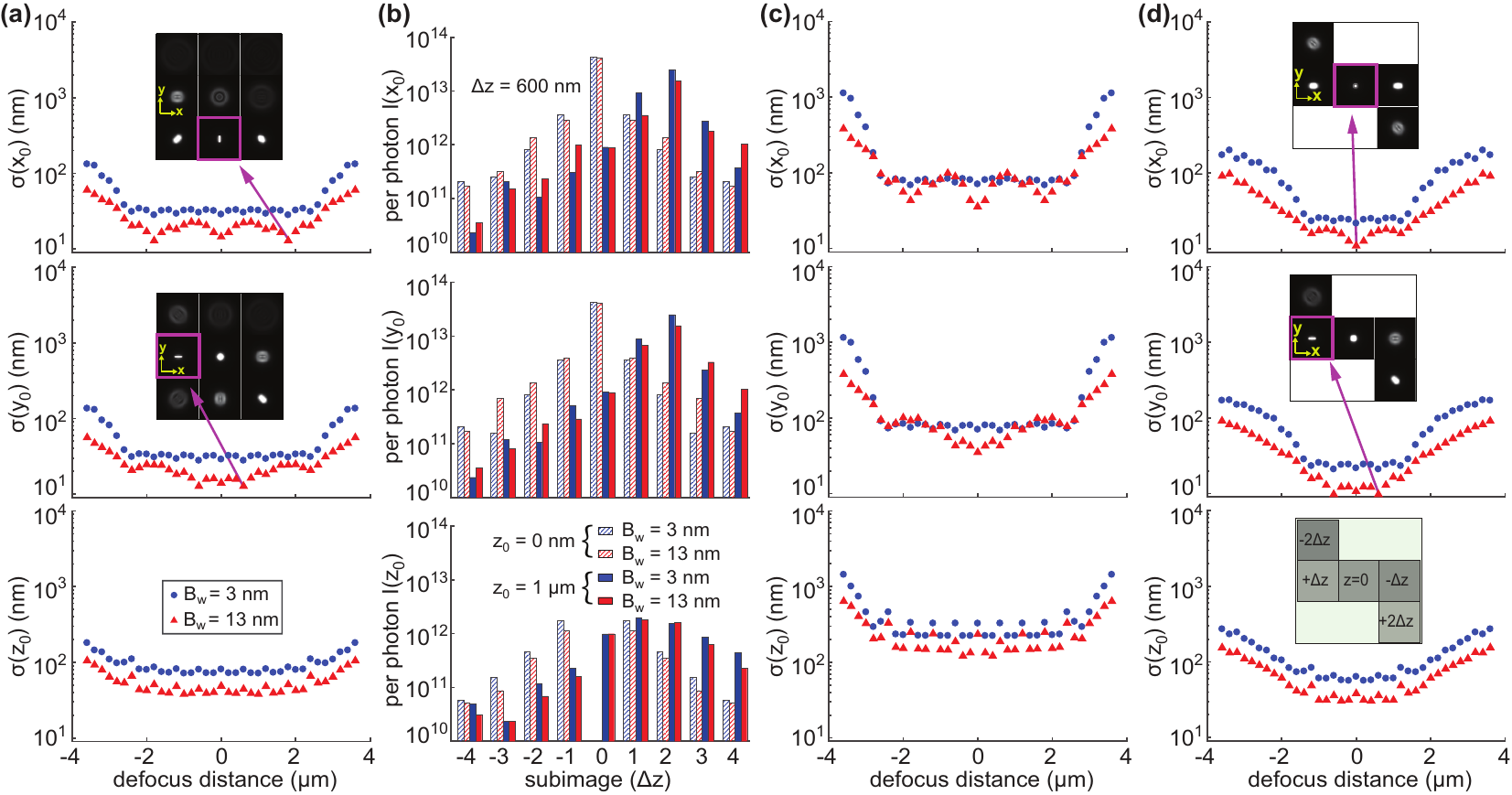}
	\caption{Numerical results of Cram\'er-Rao lower bound (CRLB) and Fisher information for multifocal systems. (a) CRLB of a point emitter imaged in a noise-free 9-plane chromatic multifocal microscope. The inset in top panel represents the $B_\textrm{w}$= 13\,nm multifocal image for an emitter at defocus distance of 1800 nm with the in-focus subimage boxed by magenta. Similarly, the mid-panel inset is for a defocus distance of 600\,nm. (b) The single-photon Fisher information for each multifocal subimage. Notice the zero single-photon Fisher information when the emitter is in focus with $z_0=0$. (c) CRLB of a point emitter imaged in a 9-plane chromatic multifocal microscope with background and measurement noise. (d) Similar to (c) but for a 5-plane chromatic multifocal microscope, where the bottom-panel inset illustrates this 5-plane arrangement. The top-panel inset shows the multifocal image for an emitter at 0-nm defocus distance, while the mid-panel inset shows the multifocal image for an emitter at 600-nm defocus distance. For these calculations, a 15 $\times$ 15 pixels$^2$ detector with a pixel size of 8\,\textmu m is used with $x_0=0$, $y_0=0$, $\lambda_\textrm{c}=520$\,nm, $M=100$, $\mathit{NA}=1.4$, $n_\textrm{oil}=1.515$, $t=100$\,ms, $f_4=200$\,mm, $d_{\textrm{g}}=32$\,\textmu m, and $\Re^{2}$ spans $300\times300$\,\textmu m$^{2}$. $N$ for $B_\textrm{w}=13$\,nm is empirically found to be 4 times more versus $B_\textrm{w}=3$\,nm. Assuming $N=500$\,counts for the $B_\textrm{w}=13$\,nm system, we use $N=125$\,counts for $B_\textrm{w}=3$\,nm simulation. The oscillatory feature in the CRLB curves is due to the integrands in calculating the Fisher information. The combination of Bessel functions in $(x,y)$ and trigonometric functions in $z_0$, coupled with the finite camera region, becomes increasingly oscillatory with increasing defocus distance. The noise parameters used in (c--d) are $\beta(k,t)=20$\,photons/pixel, $\eta_{k}=0$ and $\sigma_{\omega,k}=6$\,$\emph{e}^{-1}$/\textrm{pixel rms} (see Appendix for noise model).}
	\label{fig:myfig2}\hrule
\end{figure*}

	The numerical results of $\sigma$ for a 9-plane chromatic multifocal microscope are displayed in Figs.~\ref{fig:myfig2}a. These plots show improved localization precision (smaller $\sigma$) in all three dimensions using $B_\text{w}=13$\,nm versus $B_\text{w}=3$\,nm. Although the axial $\sigma(z_0)$ is expected to be different from the lateral $\sigma$'s, $\sigma(x_0)$ and $\sigma(y_0)$ are also different from each other across the emitter defocus range due to the varying chromatic dispersion in $x$ and $y$ directions across the multifocal subimages. For $B_\text{w} = 13$\,nm in particular, a minima for $\sigma(x_0)$ occurs at a defocus distance of 1800\,nm coinciding with the emitter being in focus at the $+3\Delta z$ subimage. The corresponding multifocal image for this point, displayed in the top-panel inset of Fig.~\ref{fig:myfig2}a, shows no dispersion in the $x$ direction for this in-focus subimage leading the low $\sigma(x_0)$. The same reasoning explains the minima locations of the $\sigma(y_0)$, $B_\text{w} = 13$\,nm data where for a defocus distance of 600\,nm, there is no dispersion in the $y$ direction in the $+\Delta z$ subimage indicated by the magenta box in the mid-panel inset. Therefore, the varying PSF asymmetry arising from chromatic dispersion affects the 3D localization precision, differently in each axis. 
	
	To investigate how the multifocal subimages contribute to the overall localization precision, we study the information content carried by each detected photon (Fig.~\ref{fig:myfig2}b) since individual photons are the most basic information carrier in single-molecule studies \cite{watkins_yang_2004}. The Fisher information is maximum around subimages where the emitter is in-focus, but it decreases as the emitter becomes more defocused. The information content per photon for estimating $z_0$ is seen to be consistently lower than those for estimating $x_0$ and $y_0$, not surprisingly, to result in an overall lower localization precision along the axial direction. Although the single-photon Fisher information for $B_\text{w} = 3$\,nm and $13$\,nm could be greater or smaller relative to each other across subimages, their cumulative effect seen in the corresponding CRLB plots (Fig.~\ref{fig:myfig2}a) consistently show lower $\sigma$ values for the $B_\text{w}=13$\,nm due to its greater photon throughput. 
	
	In the presence of noise, the localization precision degrades as expected (Fig.~\ref{fig:myfig2}c). It is important to note, however, that the quantitative values of localization precision depend significantly on the noise model and the parameters used in the model; for practical applications, one may construct an experimentally realistic noise model which our framework allows. When fewer multifocal subimages are used, e.g., a 5-plane multifocal microscope, the localization precision (Fig.~\ref{fig:myfig2}d) is seen to be consistently better that of a 9-plane microscope (cf.\ Fig.~\ref{fig:myfig2}c). This is because in a 5-plane microscope, emission photons are divided into 5 planes to result in increased number of per-subimage photons. On the other hand, for the same $\Delta z$ subimage offset, a 9-plane system affords a greater dynamic  particle defocus range ($(9-1)\times 600=4800$\,nm) compared to a 5-plane system ($(5-1)\times 600=2400$\,nm).
	
	
	Our results indicate that wider bandwidth emission filters having higher signal throughput provide better localization. This is true irrespective of the emitted photon number. Thus our analysis is directly applicable to single-molecule localization applications such as multifocal-based Stochastic Optical Reconstruction Microscopy (STORM)  \cite{Oudjedi:16}. The increased photon throughput also enables higher time resolution volume imaging data acquisition for investigating fast dynamics, without needing customized chromatic correction optics. It ought be noted, however, that localization precision depends on other factors as well. In dense particle populations, for example, broadening of the chromatic PSF with increased bandwidth could result in overlap of single molecules in the image, leading to signal crosstalk which reduces localization precision. Note that for imaging applications involving larger structures, chromatic dispersion decreases image quality. In such cases, post-processing methods including image deconvolution using the measured PSF for each subimage is one way to minimize this limitation \cite{sarder2006deconvolution}. It remains to be explored how such image deconvolution techniques are impacted by emission filter bandwidth. Furthermore, note that for the chromatic PSF modeling in these simulations, we assumed a constant emission intensity profile across the wavelengths within the bandwidth values considered. For larger bandwidth analysis, the varying emission profile across the wavelengths within a given bandwidth should be taken into account in the PSF modeling step of the simulations for precise localization results. Future work also involves modeling aberrations originating from other optics in the system.
	
	\section{Conclusion}
	In summary, this paper presents a theoretical framework to investigate the effect of chromatic dispersion arising from multifocal gratings, in multifocal microscopes, on the 3D localization precision. Simulation results show that increased filter bandwidth improves the localization in the lateral and axial position estimation despite the increased PSF distortion due to the grating induced chromatic dispersion. This indicates that the increased photon throughput at higher emission filter bandwidths has a larger effect on the localization precision, compared to the distorted image function. This study serves as a guide for researchers to help optimize existing chromatic multifocal microscopes, making these instruments increasingly useful for dynamical imaging applications. 
	
	\vspace{0.5em}
	\section*{Funding}
	The Princeton University Eric and Wendy Schmidt Transformative Technology Fund.
	
	\vspace{0.5em}
	\section*{Disclosures} 
	The authors declare no conflicts of interest.
	
		\bibliography{slmtheoryrefs}

\begin{thebibliography}{10}
\urlstyle{rm}
\expandafter\ifx\csname url\endcsname\relax
  \def\url#1{\texttt{#1}}\fi
\expandafter\ifx\csname urlprefix\endcsname\relax\def\urlprefix{URL }\fi
\expandafter\ifx\csname doiprefix\endcsname\relax\def\doiprefix{DOI: }\fi
\providecommand{\bibinfo}[2]{#2}
\providecommand{\eprint}[2][]{\url{#2}}

\bibitem{dalgarno2010multiplane}
\bibinfo{author}{Dalgarno, P.~A.} \emph{et~al.}
\newblock \bibinfo{journal}{\bibinfo{title}{Multiplane imaging and three
  dimensional nanoscale particle tracking in biological microscopy}}.
\newblock {\emph{\JournalTitle{Opt. Express}}} \textbf{\bibinfo{volume}{18}},
  \bibinfo{pages}{877--884} (\bibinfo{year}{2010}).

\bibitem{abrahamsson2013fast}
\bibinfo{author}{Abrahamsson, S.} \emph{et~al.}
\newblock \bibinfo{journal}{\bibinfo{title}{Fast multicolor 3d imaging using
  aberration-corrected multifocus microscopy}}.
\newblock {\emph{\JournalTitle{Nat. Methods}}} \textbf{\bibinfo{volume}{10}},
  \bibinfo{pages}{60} (\bibinfo{year}{2013}).

\bibitem{von2017three}
\bibinfo{author}{von Diezmann, A.}, \bibinfo{author}{Shechtman, Y.} \&
  \bibinfo{author}{Moerner, W.}
\newblock \bibinfo{journal}{\bibinfo{title}{Three-dimensional localization of
  single molecules for super-resolution imaging and single-particle tracking}}.
\newblock {\emph{\JournalTitle{Chemical reviews}}}
  \textbf{\bibinfo{volume}{117}}, \bibinfo{pages}{7244--7275}
  (\bibinfo{year}{2017}).

\bibitem{nehme2020deepstorm3d}
\bibinfo{author}{Nehme, E.} \emph{et~al.}
\newblock \bibinfo{journal}{\bibinfo{title}{Deepstorm3d: dense 3d localization
  microscopy and psf design by deep learning}}.
\newblock {\emph{\JournalTitle{Nature Methods}}} \textbf{\bibinfo{volume}{17}},
  \bibinfo{pages}{734--740} (\bibinfo{year}{2020}).

\bibitem{ma2016three}
\bibinfo{author}{Ma, Q.} \emph{et~al.}
\newblock \bibinfo{journal}{\bibinfo{title}{Three-dimensional fluorescent
  microscopy via simultaneous illumination and detection at multiple planes}}.
\newblock {\emph{\JournalTitle{Sci. Rep.}}} \textbf{\bibinfo{volume}{6}},
  \bibinfo{pages}{1--8} (\bibinfo{year}{2016}).

\bibitem{yoo2018bayesian}
\bibinfo{author}{Yoo, S.} \emph{et~al.}
\newblock \bibinfo{journal}{\bibinfo{title}{Bayesian approach for automatic
  joint parameter estimation in 3d image reconstruction from multi-focus
  microscope}}.
\newblock {\emph{\JournalTitle{25th IEEE International Conference on Image
  Processing (ICIP)}}} \bibinfo{pages}{3583--3587} (\bibinfo{year}{2018}).

\bibitem{walker2019high}
\bibinfo{author}{Walker, B.~J.} \& \bibinfo{author}{Wheeler, R.~J.}
\newblock \bibinfo{journal}{\bibinfo{title}{High-speed multifocal plane
  fluorescence microscopy for three-dimensional visualisation of beating
  flagella}}.
\newblock {\emph{\JournalTitle{J. Cell Sci.}}} \textbf{\bibinfo{volume}{132}},
  \bibinfo{pages}{jcs231795} (\bibinfo{year}{2019}).

\bibitem{lin2019multi}
\bibinfo{author}{Lin, W.}, \bibinfo{author}{Wang, D.}, \bibinfo{author}{Meng,
  Y.} \& \bibinfo{author}{Chen, S.-C.}
\newblock \bibinfo{journal}{\bibinfo{title}{Multi-focus microscope with hilo
  algorithm for fast 3-d fluorescent imaging}}.
\newblock {\emph{\JournalTitle{PloS One}}} \textbf{\bibinfo{volume}{14}}
  (\bibinfo{year}{2019}).

\bibitem{aminUniform2020PLOSONE}
\bibinfo{author}{Amin, M.~J.}, \bibinfo{author}{Petry, S.},
  \bibinfo{author}{Yang, H.} \& \bibinfo{author}{Shaevitz, J.~W.}
\newblock \bibinfo{journal}{\bibinfo{title}{Uniform intensity in multifocal
  microscopy using a spatial light modulator}}.
\newblock {\emph{\JournalTitle{PLoS One}}} \textbf{\bibinfo{volume}{15}},
  \bibinfo{pages}{e0230217} (\bibinfo{year}{2020}).

\bibitem{cover2012elements}
\bibinfo{author}{Cover, T.~M.} \& \bibinfo{author}{Thomas, J.~A.}
\newblock \emph{\bibinfo{title}{Elements of information theory}}
  (\bibinfo{publisher}{John Wiley \& Sons}, \bibinfo{year}{2012}).

\bibitem{ramHow2005}
\bibinfo{author}{Ram, S.}, \bibinfo{author}{Ward, E.~S.} \&
  \bibinfo{author}{Ober, R.~J.}
\newblock \bibinfo{journal}{\bibinfo{title}{How accurately can a single
  molecule be localized in three dimensions using a fluorescence microscope?}}
\newblock {\emph{\JournalTitle{Proceedings Volume 5699, SPIE BIOS}}}
  (\bibinfo{year}{2005}).

\bibitem{tahmasbi2014designing}
\bibinfo{author}{Tahmasbi, A.} \emph{et~al.}
\newblock \bibinfo{journal}{\bibinfo{title}{Designing the focal plane spacing
  for multifocal plane microscopy}}.
\newblock {\emph{\JournalTitle{Opt. Express}}} \textbf{\bibinfo{volume}{22}},
  \bibinfo{pages}{16706--16721} (\bibinfo{year}{2014}).

\bibitem{born2013principles}
\bibinfo{author}{Born, M.} \& \bibinfo{author}{Wolf, E.}
\newblock \emph{\bibinfo{title}{Principles of optics: electromagnetic theory of
  propagation, interference and diffraction of light}}
  (\bibinfo{publisher}{Elsevier}, \bibinfo{year}{2013}).

\bibitem{ramResolution2007}
\bibinfo{author}{Ram, S.}
\newblock \emph{\bibinfo{title}{Resolution {{And Localization In Single
  Molecule Microscopy}}}}.
\newblock Ph.D. thesis, \bibinfo{school}{Biomedical Engineering},
  \bibinfo{address}{{Arlington, TX}} (\bibinfo{year}{2007}).

\bibitem{snyder2012random}
\bibinfo{author}{Snyder, D.~L.} \& \bibinfo{author}{Miller, M.~I.}
\newblock \emph{\bibinfo{title}{Random point processes in time and space}}
  (\bibinfo{publisher}{Springer Science \& Business Media},
  \bibinfo{year}{2012}).

\bibitem{watkins_yang_2004}
\bibinfo{author}{Watkins, L.~P.} \& \bibinfo{author}{Yang, H.}
\newblock \bibinfo{journal}{\bibinfo{title}{Information bounds and optimal
  analysis of dynamic single molecule measurements}}.
\newblock {\emph{\JournalTitle{Biophys. J.}}} \textbf{\bibinfo{volume}{86}},
  \bibinfo{pages}{4015--4029} (\bibinfo{year}{2004}).

\bibitem{Oudjedi:16}
\bibinfo{author}{Oudjedi, L.} \emph{et~al.}
\newblock \bibinfo{journal}{\bibinfo{title}{Astigmatic multifocus microscopy
  enables deep 3d super-resolved imaging}}.
\newblock {\emph{\JournalTitle{Biomed. Opt. Express}}}
  \textbf{\bibinfo{volume}{7}}, \bibinfo{pages}{2163--2173},
  \doiprefix\url{10.1364/BOE.7.002163} (\bibinfo{year}{2016}).

\bibitem{sarder2006deconvolution}
\bibinfo{author}{Sarder, P.} \& \bibinfo{author}{Nehorai, A.}
\newblock \bibinfo{journal}{\bibinfo{title}{Deconvolution methods for 3-d
  fluorescence microscopy images}}.
\newblock {\emph{\JournalTitle{IEEE Signal Processing Magazine}}}
  \textbf{\bibinfo{volume}{23}}, \bibinfo{pages}{32--45}
  (\bibinfo{year}{2006}).

\end{thebibliography}
	
	\vspace{0.5em}
	\section*{Appendix}
	 To account for noise due to background and detector readout, the Fisher information matrix elements in \eqref{eq:Fisher_no_gaussian_noise} is modified to \cite{ramHow2005},
	$$
	I_{ij}^{\textrm{noise}}(\theta)=\sum_{k=1}^{P}\left(\dfrac{\partial\mu_{\theta}(k,t)}{\partial\theta_i} \right)\left(\dfrac{\partial\mu_{\theta}(k,t)}{\partial\theta_j} \right)\left[\zeta(k,t)-1\right].
	$$
	Here, $\zeta(k,t)$ is expressed as,
	\begin{equation}\label{eq:NoiseTerm_supp}
	\zeta(k,t)=\int_{\Re}\dfrac{\left(\sum_{l=1}^{\infty}\dfrac{[v_{\theta}(k,t)]^{l-1}e^{-v_{\theta}(k,t)}}{(l-1)!}\dfrac{e^{-\dfrac{1}{2}\left(\dfrac{z-l-\eta_{k}}{\sigma_{\omega,k}}\right)^{2}}}{\sqrt{2\pi}\sigma_{\omega,k}}\right)^{2}}{\dfrac{1}{\sqrt{2\pi}\sigma_{\omega,k}}\sum_{l=0}^{\infty}\dfrac{[v_{\theta}(k,t)]^{l}e^{-v_{\theta}(k,t)}}{l!}e^{-\dfrac{1}{2}\left(\dfrac{z-l-\eta_{k}}{\sigma_{\omega,k}}\right)^{2}}}dz,
	\end{equation}
	where $v_{\theta}(k,t) = \mu_{\theta}(k,t) + \beta(k,t), k = 1, ..., P$. 
	



\end{document}